# Dynamics of upstream flame propagation in a hydrogen-enriched premixed flame

Ashoke De

Department of Aerospace Engineering, Indian Institute of Technology Kanpur, Kanpur, 208016, India

Sumanta Acharya*

Turbine Innovation and Energy Research Center, Louisiana State University, Baton Rouge, LA 70803, USA

**Abstract:** An unconfined strongly swirled flow is investigated to study the effect of hydrogen addition on upstream flame propagation in a methane-air premixed flame using Large Eddy Simulation (LES) with a Thickened Flame (TF) model. A laboratory-scale swirled premixed combustor operated under atmospheric conditions for which experimental data for validation is available has been chosen for the numerical study. In the LES-TF approach, the flame front is resolved on the computational grid through artificial thickening and the individual species transport equations are directly solved with the reaction rates specified using Arrhenius chemistry. Good agreement is found when comparing predictions with the published experimental data including the predicted RMS fluctuations. Also, the results show that the initiation of upstream flame propagation is associated with balanced maintained between hydrodynamics and reaction. This process is associated with the upstream propagation of the center recirculation bubble, which pushes the flame front in the upstream mixing tube. Once the upstream movement of the flame front is initiated, the hydrogen-enriched mixture exhibits more unstable behavior; while in contrast, the $CH_4$ flame shows stable behavior.

**Keywords**: Premixed hydrogen combustion, Gas turbine, Large Eddy Simulation, Thickened-Flame

*Corresponding Author:

E-mail address: acharya@tigers.lsu.edu

## 1. INTRODUCTION

Recent interest in clean coal technology and the associated development of the Integrated Gasification Combined Cycle (IGCC) for power generation has led to studies dealing with the combustion of coal-gasification products commonly referred to as syngas. Hydrogen ($H_2$) is a primary component of syngas with the overall composition contains varying fractions of CO, $CH_4$ and other hydrocarbons which are largely dependent on the gasification process. The presence of hydrogen in syngas plays a significant role in the flame behavior given its much higher flame speed (nearly 5 times that of methane) and molecular diffusivity (Lewis number of nearly 0.3 compared to 1 for methane). Even for premixed methane flames, it has been suggested that hydrogen addition can improve flame holding and Lean Blow Out (LBO) behavior. Therefore, it is desirable to understand how mixing of natural gas (or methane) and syngas (or hydrogen) in different proportions effects the combustion characteristics. Such understanding is also important for syngas combustor operations in order to handle the variabilities in fuel composition without adverse performance issues [1]. This is a primary reason why fuel flexibility is an important consideration for combustor design. The major operability issues associated with lean premixed combustions are: flame-holding, flashback, combustion instabilities and auto-ignition.

In order to achieve stable combustion, hydrogen doping appears to be a promising approach toward increasing LBO limits of hydrocarbon fuels [2]. While increase in hydrogen proportion can potentially improve flame stability, at the same time, it can make the combustor more susceptible to flashback (upstream flame propagation) and thermo-acoustic instability. Thus, the role of hydrogen on the flame behavior needs to be carefully examined and understood.

Flashback becomes an inherent problem with hydrogen since hydrogen flame speeds are quite high [3]. Flashback has been investigated by several researchers [4-6] and it is reported that the behavior becomes substantially more complex in swirling flows [7]. There are several mechanisms reported in the literature [8-10], but the occurrence of flashback is unique to each system. Different flashback mechanisms have already been reported in literature as: (i) Flashback via flame propagation [5, 11], (ii) Flashback due to combustion instabilities, (iii) Combustion-induced vortex breakdown [8, 10-19].

Most land-based gas turbine systems operate in a swirl, lean premixed mode, and are particularly susceptible to flashback with hydrogen or syngas. Flashback into the premixer section leads to thermal overload and destruction of the hardware therefore it must be avoided at all load conditions [10]. Flashback can be prevented by using specially designed flame holders or by injecting syngas in a separate non-premixed arrangement. However, in transitioning from natural gas as the fuel of choice to syngas, it is desirable to keep hardware changes to a minimum, given the extensive body of knowledge with current natural gas related hardware. Therefore, a fundamental understanding of the flame behavior in swirled premixed systems with hydrogen addition is needed, and serves as the motivation for the present study.

The configuration of interest in the present work is hydrogen-enriched swirl-stabilized flame. Several studies have investigated premixed flames of $H_2$-hydrocarbon fuel mixtures.

Recent examples include studies by Morris et al. [20] in a commercial gas turbine engine operating in the LPM combustion mode, Schefer [21], Schefer et al. [22], Kim et al. [23-24], Bellester et al. [25] for swirling hydrogen-enriched flames, Griebel et al [2] for hydrogen-enriched flames at higher pressure, and Strakey et al. [26] for swirl stabilized hydrogen-enriched flames at high pressure. Kröner et al. [10, 14] also reported flashback due to combustion induced vortex breakdown (CIVB) in swirling flows for different $CH_4+H_2$ mixtures. They reported that the CIVB happened to be the prevailing mechanism in a swirled burner without center-body. More recently, Hegger et al. [18] and Tangermann et al. [19] studied the same burner to analyze the flashback behavior using both experimental and computational technique. Tuncer et al. [27] also investigated flashback characteristics in a confined premixed hydrogen-enriched methane flames for a laboratory scale swirled combustor. An extensive review on swirling flows for combustion systems can be found in the literature [28-32].

In this investigation, we study flame behavior in hydrogen enriched premixed swirling flames using LES combined with a TF approach for combustion. Specific goal of this study is to understand the flame behavior in swirling flames, and in particular, how hydrogen enrichment influences this particular aspect.

## 2. FLOW CONFIGURATION

The configuration considered here is an unconfined swirl burner as shown in Fig. 1, and corresponds to a configuration for which measurements of the velocity field in a reacting flow is available [30] for validation of the numerical simulations. The 45° swirl vane is fitted with a solid center body which also acts as a fuel injector [27, 30]. This center body extends beyond the swirl vane and is flush with the dump plane of the combustor. The diameter of the center body is 12.7mm (0.5 inch) and the outer diameter (O.D.) of the swirler is 34.9 mm (1.375 inch). Methane and hydrogen gas is injected radially from the center body through eight holes immediately downstream of the swirler vane. The fuel/air mixer is assumed to be premixed at the dump plane and the equivalence ratio is calculated to be φ=0.7 using the following relationship.

$$\Phi_{CH4+H2} = \left[ (X_{CH4} + X_{H2}) / X_{O2} \right] / \left[ (X_{CH4} + X_{H2}) / X_{O2} \right]_{stoich} \tag{1}$$

The Reynolds number Re is based on the inlet bulk velocity and hydraulic diameter, and the geometric swirl number $S_g$ is defined as the ratio of the axial flux of the tangential momentum to the product of axial momentum flux and a characteristic radius ,

$$S_g = \frac{1}{R} \frac{\int_0^R r^2 \overline{U}_x \overline{W} dr}{\int_0^R r \overline{U}_x^2 dr} \tag{2}$$

The investigation in this paper is carried out for Reynolds number Re=10144 and Swirl number $S_g$=0.82 at atmospheric pressure and temperature.

## 3. NUMERICAL DETAILS

In turbulent premixed combustion, a popular approach is to rely on the flamelet concept, which essentially assumes the reaction layer thickness to be smaller than the smallest turbulence scales. The two most popular model based on this concept are the flame surface density model (FSD) [33] and the G-equation model [34-35]. It has been reported that the FSD model is not adequate beyond the corrugated flamelet regime [36-37], while the G-equation approach depends on a calculated signed-distance function that represents an inherent drawback of this method. Another family of models relies on the probability density function (PDF) approach [38], which directly considers the probability distribution of the relevant quantities in a turbulent reacting flow. Moreover, it can be applied to non-premixed, premixed, and partially premixed flames without having much difficulty. Usually, there are two ways which are mainly used to calculate the pdf: one is *presumed pdf approach*, and other is *pdf transport balance equation approach*. The *presumed pdf approach*, which essentially assumes the shape of the probability function P, is relatively simpler to use, however, has severe limitations in the context of applicability. On the other hand, the *pdf transport balance equation approach* solves a transport equation for pdf function, which is applicable for multi species, mass-weighted probability density function. This method has considerable advantage over any other turbulent combustion model due to its inherent capability of handling any complex reaction mechanism. However, the major drawback of transport pdf approach is its high dimensionality, which essentially makes the implementation of this approach to different numerical techniques, like FVM or FEM, limited, since their memory requirements increase almost exponentially with dimensionality. Usually, Monte-Carlo algorithms, which reduce the memory requirements, are used by Pope [38]. Moreover, a large number of particles need to be present in each grid cell to reduce the statistical error; however this makes it a very time consuming process. So far, the transport equation method has been only applied to relatively simple situations.

In this work, a Thickened Flame (TF) model [39] is invoked where the flame is artificially thickened to resolve it on computational grid points where reaction rates from kinetic models are specified using reduced mechanisms. The influence of turbulence is represented by a parameterized efficiency function. A key advantage of the TF model is that it directly solves the species transport equations and uses the Arrhenius formulation for the evaluation of the reaction rates. Another major advantage associated with this TF model is the ability to capture the complex swirl stabilized flame behavior with fully or partially premixed inlet reactants which are often found in a gas turbine combustor. The present TF model is capable of taking care of partially premixed gas since for the solution is obtained for the individual species transport equations and the reaction rates are specified using Arrhenius expressions.

## 3.1 Governing equations and flow modeling using LES

The filtered governing equations for the conservation of mass, momentum, energy and species transport are given as:

Continuity equation:

$$\frac{\partial}{\partial t}(\bar{\rho}) + \frac{\partial}{\partial x_i}(\bar{\rho}\,\bar{u_i}) = 0 \tag{3}$$

Momentum equation:

$$\frac{\partial}{\partial t}(\bar{\rho}\,\bar{u_i}) + \frac{\partial}{\partial x_j}(\bar{\rho}\,\bar{u_i}\,\bar{u}_j) = -\frac{\partial}{\partial x_i}(\bar{p}) + \frac{\partial}{\partial x_j}\left((\mu+\mu_t)\frac{\partial \bar{u_i}}{\partial x_j}\right) \tag{4}$$

Energy equation:

$$\frac{\partial}{\partial t}(\bar{\rho}\bar{E}) + \frac{\partial}{\partial x_i}(\bar{\rho}\bar{u_i}\bar{E}) = -\frac{\partial}{\partial x_j}\left(\bar{u}_j\left(-\bar{p}I + \mu\frac{\partial \bar{u_i}}{\partial x_j}\right)\right) + \frac{\partial}{\partial x_i}\left(\left(k+\frac{\mu_t C_p}{\mathrm{Pr}_t}\right)\frac{\partial \bar{T}}{\partial x_i}\right)$$
$$+ \frac{\partial}{\partial x_i}\left(\bar{\rho}\sum_{s=1}^{N} h_s\left(D+\frac{\mu_t}{Sc_t}\right)\frac{\partial \bar{Y_s}}{\partial x_i}\right) \tag{5}$$

Species transport equation:

$$\frac{\partial}{\partial t}(\bar{\rho}\bar{Y_i}) + \frac{\partial}{\partial x_j}(\bar{u}_j\,\bar{\rho}\,\bar{Y_i}) = \frac{\partial}{\partial x_j}\left(\left(D+\frac{\mu_t}{Sc_t}\right)\frac{\partial \bar{Y_i}}{\partial x_j}\right) + \overline{\dot{\omega}_i} \tag{6}$$

where ρ is the density, $u_i$ is the velocity vector, p is the pressure, $E = e + u^2_i/2$ is the total energy, (where e = h - p/ρ is the internal energy and h is enthalpy), $h_f^0$ is enthalpy of formation, and $\dot{\omega}$ is the reaction rate. The fluid properties μ, $k$ and $D$ are respectively the viscosity, the thermal conductivity and molecular diffusivity, while $\mu_t$, $Sc_t$ and $Pr_t$ are the turbulent eddy viscosity, the turbulent Schmidt number and the turbulent Prandtl number respectively.

To model the turbulent eddy viscosity, LES is used so that the energetic larger-scale motions are resolved, and only the small scale fluctuations are modeled. The sub-grid stress modeling is done using a dynamic Smagorinsky model where the unresolved stresses are related to the resolved velocity field through a gradient approximation:

$$\overline{u_i u_j} - \overline{u_i}\,\overline{u_j} = -2\nu_t \overline{S}_{ij} \tag{7}$$

where
$$\nu_t = C_s{}^2 (\Delta)^2 \left| \overline{S} \right| \tag{8}$$

$$\overline{S}_{ik} = \frac{1}{2}\left( \frac{\partial \overline{u_i}}{\partial x_k} + \frac{\partial \overline{u_k}}{\partial x_i} \right) \tag{9}$$

$$\left| \overline{S} \right| = \sqrt{2\overline{S}_{ik}\,\overline{S}_{ik}} \tag{10}$$

and S is the mean rate of strain. The coefficient $C_s$ is evaluated dynamically [40-41] and locally-averaged.

## 3.2 Thickened-Flame (TF) Modeling Approach with LES:

As noted earlier, we have used the TF modeling technique, where the flame front is artificially thickened to resolve on computational grid. Corrections are made to ensure that the flame is propagating at the same speed as the un-thickened flame [39, 42-43]. The key benefit of this approach, as noted earlier, rests in the ability to computationally resolve the reaction regions and the chemistry in these regions.

Butler and O'Rourke [44] were the first to propose the idea of capturing a propagating premixed flame on a coarser grid. The basic idea with this approach is that the flame is artificially thickened to include several computational cells and by adjusting the diffusivity to maintain the same laminar flame speed $s_L^0$. It is well known from the simple theories of laminar premixed flames [34, 45] that the flame speed and flame thickness can be related through the following relationship

$$s_L^0 \propto \sqrt{D\overline{B}}, \delta_L^0 \propto \frac{D}{s_L^0} = \sqrt{\frac{D}{\overline{B}}} \tag{11}$$

where D is the molecular diffusivity and $\overline{B}$ is the mean reaction rate. When the flame thickness is increased by a factor F, the molecular diffusivity and reaction rate are modified accordingly (FD and $\overline{B}$/F) to maintain the same flame speed. The major advantages associated with thickened flame modeling are: (i) the thickened flame front is resolved on LES mesh which is usually larger than typical premixed flame thickness (around 0.1-1 mm), (ii) quenching and ignition events can be simulated, (iii) chemical reaction rates are calculated exactly like in a DNS calculation without any *ad-hoc* sub models, so it can theoretically be extended to incorporate with multi-step chemistry [39].

In LES framework, the spatially filtered species transport equation is given in Equation (6), where the terms on the right hand side are the filtered diffusion flux plus the unresolved transport, and the filtered reaction rate respectively. In general, the unresolved term is modeled with a gradient diffusion assumption by which the laminar diffusivity is augmented by the turbulent eddy diffusivity. However, in the TF model, the "thickening" procedure multiplies the diffusivity term by a factor F which has the effect of augmenting the diffusivity. Therefore, the gradient approximation for the unresolved fluxes is not explicitly used in the closed species transport equations. The corresponding filtered species transport equation in the thickened-flame model becomes

$$\frac{\partial \overline{\rho Y_i}}{\partial t} + \frac{\partial}{\partial x_j}(\overline{\rho Y_i u_j}) = \frac{\partial}{\partial x_j}\left(\overline{\rho} F D_i \frac{\partial \overline{Y_i}}{\partial x_j}\right) + \frac{\overline{\dot{\omega}_i}}{F} \tag{12}$$

Although the filtered thickened flame approach looks promising, a number of key issues need to be addressed. The thickening of the flame by a factor of F modifies the interaction between turbulence and chemistry, represented by the Damköhler number, Da, which is a ratio of the turbulent ($\tau_t$) and chemical ($\tau_c$) time scales. Da, is decreased by a factor F and becomes Da/F, where

$$Da = \frac{\tau_t}{\tau_c} = \frac{l_t s_L^0}{u' \delta_L^0} \qquad (13)$$

As the *Da* is decreased, the thickened flame becomes less sensitive to turbulent motions. Therefore, the sub-grid scale effects have been incorporated into the thickened flame model, and parameterized using an efficiency function E derived from DNS results [39]. Using the efficiency function, the final form of species transport equation becomes

$$\frac{\partial \overline{\rho Y_i}}{\partial t} + \frac{\partial}{\partial x_j}(\overline{\rho Y_i u_j}) = \frac{\partial}{\partial x_j}\left(\overline{\rho} E F D_i \frac{\partial \overline{Y_i}}{\partial x_j}\right) + E \frac{\overline{\dot{\omega}_i}}{F} \qquad (14)$$

where the modified diffusivity ED, before multiplication by F to thicken the flame front, may be decomposed as ED=D(E-1)+D and corresponds to the sum of molecular diffusivity, D, and a turbulent sub-grid scale diffusivity, (E-1)D. In fact, (E-1)D can be regarded as a turbulent diffusivity used to close the unresolved scalar transport term in the filtered equation.

The central ingredient of the TF model is the sub-grid scale wrinkling function E, which is defined by introducing a dimensionless wrinkling factor Ξ. The factor Ξ is the ratio of flame surface to its projection in the direction of propagation. The efficiency function, E, is written as a function of the local filter size ($\Delta_e$), local sub-grid scale turbulent velocity ($u'_{\Delta_e}$), laminar flame speed ($s_L^0$), and the thickness of the laminar and the artificially thickened flame ($\delta_L^0, \delta_L^1$). Colin et al. [39] proposed the following expressions for modeling the efficiency function.

$$\Xi = 1 + \beta \frac{u'_{\Delta_e}}{s_L^0} \Gamma \left( \frac{\Delta_e}{\delta_L^0}, \frac{u'_{\Delta_e}}{s_L^0} \right) \tag{15}$$

$$\beta = \frac{2\ln 2}{3c_{ms}\left( \mathrm{Re}_t^{1/2} - 1 \right)}, c_{ms} = 0.28, \mathrm{Re}_t = u'l_t / \nu \tag{16}$$

where $\mathrm{Re}_t$ is the turbulent Reynolds number. The local filter size $\Delta_e$ is related with laminar flame thickness as

$$\Delta_e = \delta_L^1 = F\delta_L^0 \tag{17}$$

The function $\Gamma$ represents the integration of the effective strain rate induced by all scales affected due to artificial thickening, $\Gamma$ is estimated as

$$\Gamma \left( \frac{\Delta_e}{\delta_L^0}, \frac{u'_{\Delta_e}}{s_L^0} \right) = 0.75 \exp \left[ -1.2 \left( \frac{u'_{\Delta_e}}{s_L^0} \right)^{-0.3} \right] \left( \frac{\Delta_e}{\delta_L^0} \right)^{2/3} \tag{18}$$

Finally, the efficiency function takes the following form as defined by the ratio between the wrinkling factor, $\Xi$, of laminar flame ($_{\delta_L = \delta_L^0}$) to thickened flame ($_{\delta_L = \delta_L^1}$).

$$E = \frac{\Xi \big|_{\delta_L = \delta_L^0}}{\Xi \big|_{\delta_L = \delta_L^1}} \geq 1 \tag{19}$$

where the sub-grid scale turbulent velocity is evaluated as $u'_{\Delta_e} = 2\Delta_x^3 \left| \nabla^2 \left( \nabla \times \overline{u} \right) \right|$, and $\Delta_x$ is the grid size. This formulation for sub-grid scale velocity estimation is free from dilatation. Usually, $\Delta_e$ differs from $\Delta_x$, and it has been suggested that values for $\Delta_e$ be at least $10\Delta_x$ [39].

There are different versions of TF model available in literature depending on the calculation of E such as: Power-law flame wrinkling model [46-47], Dynamically modified TF model [43, 48]. In one of the previous studies by the authors reported that these different versions of TF models do not exhibit substantial differences in flow filed and temperature predictions [49]. That's why we use the original formulation of TF model for the present study [39]. More detailed description of E can found in other literature [50].

### 3.3 Chemistry model

As all the species are explicitly resolved on the computational grid, the TF model is best suited to resolve major species. Intermediate radicals with very short time scales can not be resolved. To this end, only simple global chemistry has been used with the thickened flame model.

For $CH_4$ combustion, a two step chemistry, which includes six species ($CH_4$, $O_2$, $H_2O$, $CO_2$, $CO$ and $N_2$) is used and given by the following equation set.

$$CH_4+1.5O_2 \rightarrow CO+2H_2O \tag{20}$$

$$CO+0.5O_2 \leftrightarrow CO_2 \tag{21}$$

To incorporate $H_2$ reaction in addition to the above $CH_4$ chemistry, the following 1-step Marinov mechanism is employed.

$$H_2+0.5O_2 \rightarrow H_2O \tag{22}$$

The corresponding reaction rate expressions are given by:

$$q_1=A_1 exp(-E^1_a/RT)[CH_4]^{a1}[O_2]^{b1} \tag{23}$$

$$q_2(f)=A_2 exp(-E^2_a/RT)[CO][O_2]^{b2} \tag{24}$$

$$q_2(b)=A_2 exp(-E^2_a/RT)[CO_2] \tag{25}$$

$$q_3=A_3 exp(-E^3_a/RT)[H_2][O_2]^{b3} \tag{26}$$

where the activation energy $E^1_a$ =34500 cal/mol, $E^2_a$ =12000 cal/mol, a1=0.9, b1=1.1, b2=0.5, and $A_1$ and $A_2$ are 2.e+15 and 1.e+9, as given by Selle et al. [51], and $E^3_a$ =35002 cal/mol,

b3=0.5, $A_3$=1.8e+16 (SI units) as given in the DOE report [52]. The first and third reactions (Eqs. 20 & 22) are irreversible, while the second reaction (Eq. 21) is reversible and leads to an equilibrium between CO and $CO_2$ in the burnt gases. Hence the expressions (Eqs. 23 & 26) represent the reaction rates for the irreversible reactions (Eq. 20 & 22) and the expressions (Eq. 24 & 25) represent the forward and backward reaction rates for the reversible reaction (Eq. 21). Properties including density of mixtures are calculated using CHEMKIN-II [53] and TRANFIT [54] depending on the local temperature and the composition of the mixtures at 1 atm.

### 3.4 Solution procedure

In the present study, a parallel multi-block compressible flow code for an arbitrary number of reacting species, in generalized curvilinear coordinates, is used. Chemical mechanisms and thermodynamic property information of individual species are input in standard Chemkin format. Species equations along with momentum and energy equation are solved implicitly in a fully coupled fashion using a low Mach number preconditioning technique, which is used to effectively rescale the acoustics scale to match that of convective scales [55]. An Euler differencing for the pseudo time derivative and second order backward 3-point differencing for physical time derivatives are used. A second order low diffusion flux-splitting algorithm is used for convective terms [56]. However, the viscous terms are discretized using second order central differences. An incomplete Lower-Upper (ILU) matrix decomposition solver is used. Domain decomposition and load balancing are accomplished using a family of programs for partitioning unstructured graphs and hypergraphs and computing fill-reducing orderings of sparse matrices, METIS. The message communication in distributed computing environments is achieved using Message Passing Interface, MPI. The multi-block structured curvilinear grids presented in this paper are generated using commercial grid generation software GridPro[TM].

### 3.5 Computational domain and boundary conditions

As noted earlier, and shown in Fig. 1, the configuration of interest in the present work is an unconfined swirled burner [30]. The computational domain extends 20D downstream of the dump plane (fuel-air nozzle exit), 13D upstream of the dump plane (location of the swirl vane exit) and 6D in the radial direction. Here, D is the center-body diameter. The finer mesh consists of 320x208x48 grid points downstream of the dump plane plus (98x32x48) + (114x22x48) grid points upstream, and contains approximately 3.94M grid points [30]. The grid resolution in the computational domain with 3.94M grid points is given as: (a) along axial direction $\Delta x/2D=0.031$ is maintained from inlet to dump plane, and then 0.03125 rest of the whole domain starting from dump-plane to outlet, (b) along the radial direction $\Delta r/2D=0.014$-0.0145 is maintained starting from centerline to lateral boundary, (c) along azimuthal direction: behind center-body $\Delta\theta/2D=0.011$-0.0327 (up to $r/2D=0.25$), in the annular shear layer $\Delta\theta/2D=0.0327$-0.089 ($r/2D=0.25$-0.6875), and finally $\Delta\theta/2D=0.089$-0.393 ($r/2D=0.6875$-3.0) where D=center-body diameter. The close-up view of the computation grid is shown in Fig. 1 where the top figure shows the block boundaries of a multi-block grid that reflects the clustering needed in the shear layer regions. More details on computation grid can be found in the literature [30].

The inflow boundary condition is assigned at the experimental location immediately downstream of the swirler blades. The mean axial velocity distribution is specified as a one-seventh power law profile to represent the fully developed turbulent pipe flow, with superimposed fluctuations at 10% intensity levels (generated using Gaussian distribution). A constant tangential velocity component is specified as determined from the swirl vane angle. Convective boundary conditions [57] are prescribed at the outflow boundary, and zero-gradient boundary conditions are applied on the lateral open boundaries. The time step used for the computation is $\Delta t=1.0e$-3 sec. where the flow through time is ~3.2sec. The fuel injection point is shown in Fig. 2, which represents a jet-in-cross flow type configuration, and the fuel-air premixing is simulated in this problem.

## 4. RESULTS AND DISCUSSION

We will first report the non-reacting LES calculations to ensure that the grid and boundary conditions are properly chosen, and to assess the cold-flow flow characteristics. This will be followed by a discussion of the reacting flow calculations where we will examine both the flow and flashback behavior in the context of $H_2$ addition.

### 4.1 Non-reacting flow results

Figure 2 shows the radial distribution of axial and tangential mean velocity profiles, and the axial and tangential velocity fluctuations at different axial locations for Re=10144. In general, the LES and the experimental data for the mean velocity profiles, and the velocity fluctuations at different axial locations are found to be in good agreement. It is observed that the shape, size, and the intensity of the recirculation zone (region of negative axial velocities at the center) are well predicted along with the overall spreading of the turbulent swirling jet. Some level of asymmetry can be observed in both the simulations and the experiments and indicate that the statistical averaging period needs to be carried out over a longer period of time. However, due to the presence of low-frequency unsteadiness in the flow, the averaging time-periods can be very large and impractical from both computational and experimental perspectives [30]. Similar observations of asymmetry in the averaged profiles have also been reported in the literature for the confined combustor geometry [58]. Also, the RMS fluctuations of the axial and tangential velocities are in good agreement with the experimental data. The peak in the axial velocity fluctuations is observed to be in the shear layer and between the location of the peak velocity and the recirculation bubble. In this region, the steepest velocity gradient $\partial U_i / \partial x_j$ is obtained and promotes the production of the peak kinetic energy. The tangential velocity fluctuations show a flatter profile than the axial velocity fluctuations and their peaks are shifted radially inwards as for the mean tangential velocities. More detailed discussion on non-reacting flow results can be found in the literature by De et al. [30].

For both the mean velocity and fluctuations, the fine mesh (3.94M grid points) results are in better agreement with the experimental data compared to those from the coarse mesh (1.22M grid points) for all the cases considered here. Hence, the fine mesh is chosen for reacting flow calculations.

## 4.2 Reacting flow results

Figure 3 shows the effect of hydrogen addition in the flow field. The axial velocity is affected due to the varying hydrogen percentage with methane fuel, and in particular, the size (length and width) and strength of the central recirculation bubble decreases with the increase in $H_2\%$ as shown in Fig. 3(a). This can be associated with the higher combustibility of hydrogen which, in turn, promotes faster reaction along the flame front and results in increased temperature in that region. Also, the observed axial velocity magnitude increases by 30-35% ($CH_4$ to $30\%H_2$) and 40-45% ($CH_4$-$50\%H_2$) along the flame front due to higher hydrogen addition to the methane fuel. The higher reaction rates result in increased temperature and thus the peak density ($\rho$) decreases ~25% ($CH_4$ to $30\%H_2$) and ~28% ($CH_4$-$50\%H_2$) in this region, increasing the axial velocity by a factor of $1/\rho$. On the other hand, the tangential velocity ($V_t$) increases only by $\rho^{-1/2}$ to maintain the radial pressure gradient in the radial momentum equation: $\rho V_t^2/r=dP/dr$. Because of this, the swirl ratio decreases by $\rho^{1/2}$ as shown in Fig. 3(a) (top figure). Hence, the overall effect is that as the flow exhibits greater acceleration along the flame front with hydrogen it leads to a smaller central-recirculation as depicted in Fig. 3(a) (bottom). Similar observations have also been reported by Kim et al. [23].

The radial distributions of mean axial velocity and axial velocity fluctuations are shown in Fig. 3(b) for 30%$H_2$ mixture. The overall agreement of the predictions with the data is found to be quite reasonable, considering the complexity of the physical processes and the configuration. With increasing axial distance the magnitude of the peak velocity decreases and the location of the peak is moved further outwards radially. While the general agreement between the data and predictions are satisfactory, and the LES results show the right qualitative features and the peak magnitudes, there are intrinsic differences. The axial velocities show a small over-prediction of the peak axial velocity particularly for r/2D>0.8. Predicted RMS fluctuations clearly exhibit two peaks [55]. The location of the peaks corresponds to the burnt and un-burnt regions in the shear layer and associated with the high velocity gradients where the turbulence production due to the mean velocity gradient is the highest. The first peak is located in the burnt region of the shear layer downstream of the center body where the temperatures are higher (Fig. 4 (a)). The second peak is observed in the un-burnt regions of the shear layer where the temperatures are relatively on the lower side. However, the measurements do not clearly show this dual-peak behavior.

Figure 4(a) shows the time-averaged temperature profiles at different hydrogen percentage added to the fuel. It is noteworthy that at higher $H_2$% the flame (represented by the region of higher temperature seen in Fig. 4(a)) moves upstream of the dump plane into the premixing section and stabilizes there. This is associated with the higher combustibility of $H_2$ that helps to promote faster reactions and flame propagation takes place. It is clear in Fig. 4(a) that for the hydrogen cases the flame moves up-to the fuel-injection point. A detailed discussion on this flame movement is provided in the following section.

### 4.2.1 Upstream flame propagation

Figure 4(b) shows the time series temperature plots for $CH_4$ and $30\%H_2$, which clearly shows how flame propagates upstream and finally stabilizes in the mixing tube, especially for $30\%H_2$ mixture. For $CH_4$, the flame front also moves slightly upstream of the dump plane but stabilizes in its vicinity as observed in Fig. 4(b). For the hydrogen case, as time advances, the flame front moves upstream all the way to the fuel injection point and stabilizes in the wake of the cross flow fuel-jet. Upstream flame propagation, usually characterized as flame flashback, occurs when the burning velocity exceeds the local flow velocity. In the present case, the flame propagation is initiated due to Combustion Induced Vortex Breakdown (CIVB) and then sustained due to favorable conditions due, in part, to the low velocity in the boundary layer and higher local burning velocity of $H_2$. CIVB is promoted by the higher gravimetric heating value and energy release that occurs with $H_2$. As the flame starts moving upstream for the $30\%H_2$, it encounters fuel rich regions due to the upstream point-fuel injection that, in turn, induces higher flame propagation velocity $H_2$.

Usually, the recirculation bubble arising from the vortex breakdown brings the heat and reactive species back to the flame tip and provides flame-holding at the dump plane. However, the recirculation zone gives rise to the azimuthal vorticity component in this region which produces positive or negative induced velocity following the Biot-Savart law, as given by Eq. 28 [13, 15] depending on the sign of azimuthal vorticity. The position and motion of this recirculation bubble largely depends on the balance maintained between the induced velocity and irrotational axial velocity. Slight changes in the flow field can alter this balance and result in the upstream movement of the recirculation bubble (shown in Fig. 6) accompanied by a higher induced negative velocity due to production of negative azimuthal vorticity. Fig. 5(a) shows the frequency spectra obtained from pressure fluctuations (Fig. 5(b)) upstream of dump plane. As observed, $CH_4$ shows a peak around 62 Hz, and both the 30%$H_2$ and 50%$H_2$ show the primary peak around 45 Hz. A small peak is also observed at 8Hz frequency and is the bulk-mode frequency ($f = c / 2\pi\sqrt{S / Vl} \sim 8.25 Hz$) of the inlet delivery tube.

As observed in Fig. 5(a) for the 30% & 50%$H_2$, the 45Hz frequency corresponds to the initial high amplitude oscillations where the upstream flame propagation is initiated due to the unsteady motion of the recirculation bubble (Fig. 6). The initial pressure fluctuations (Fig. 5(b)), for the 30% & 50%$H_2$, correspond to the complex interaction between shedding vortices past the bluff body and the motion of the recirculation bubble formed due to vortex breakdown; thereafter, the fluctuations disappear at later time instants as the flame front moves upstream (flashback occurs) [38].

Fig. 6 shows streamlines colored with temperature contours at five different time instances representing phases during a complete cycle of the observed oscillation frequency (45Hz), and the data at each phase is averaged over four of these cycles in order to eliminate some of the turbulent fluctuations arising from the higher frequency oscillations. As observed in phase1 (Fig. 6), a big recirculation bubble is already formed and another smaller vortex (shedding vortex-1) starts forming due to shedding at the bluff-body. The shedding vortex 1 is seen to extend while forming a stagnation point-1 along the bluff-body wall in the upstream section where the pressure tends to build up and causes flow reversal. At phase2 (Fig. 6), the same shedding vortex (shedding vortex-1) extends into the other side of the bluff-body forming another stagnation point-2 on that side while at the same time it grows into a bigger vortex and starts interacting with the center-recirculation bubble. Stagnation point-1 has moved further upstream as pressure reaches a peak value in between phase1 and phase2. At phase3 (Fig. 6) the shedding vortex-1 has sufficiently grown to form a central circulation bubble while another vortex (shedding vortex-2) starts forming from the saddle point and generates a new stagnation point3. At phase4 (Fig. 6) the saddle point vortex and the shedding vortex-2 merge into a single one. As pressure reaches minima in between phase3 and phase4, the forward flow pushes the saddle point further downstream and disappears in phase4. At phase5 (Fig. 6) the initiated vortex from phase4 starts growing and the cycle shows a repeatable pattern starting with phase1. This unsteady interaction at the initial stage helps to push the flame front into the upstream mixing tube, especially for 30%$H_2$ & 50 %$H_2$, and thereafter other favorable conditions promotes the flame propagation further upstream.

It should be noted that the upstream flame movement is also initiated in $CH_4$ case (~62 Hz), but it is not supported by any favorable condition that can promote flashback, and instead results in a limit cycle behavior in pressure fluctuations (Fig. 5(b)). To further illustrate this and the differences between methane and hydrogen flame behavior, Figs. 7 and 8 respectively show the flame movement in the mixing tube during flashback for the 30%$H_2$ case and for a corresponding time period for the methane case. In Fig. 7, for 30% $H_2$ the flame front is close to the dump plane at t=8s. As the flame front moves further upstream, at t=10.5s it starts interacting with the injected fuel due to an increase in pressure ahead of the flame front. The pressure difference between upstream and downstream of the flame front causes the formation of a small recirculation bubble ahead of the flame, which becomes more prominent at later time instances at t=12.0s & t=13.8s (Fig. 7). This interaction and the time-period where the flame is moving upstream in the delivery tube (flashback) can be correlated with the lower amplitude pressure fluctuations observed between 10-13s in Fig. 5(b). In contrast, the initial high amplitude fluctuations in Fig. 5(b) correspond to the dynamics shown in Fig. 6 in the early-stages of the flashback behavior where the flame is still near the dump plane and interacts with the bluff-body vortices and the swirl-induced center-recirculation (t=8s in Fig. 7).

For pure $CH_4$ (Fig. 8) the 62Hz frequency (Fig. 5(a)) corresponds to the high amplitude limit cycle behavior observed in Fig. 5(b). Here, as shown in Fig. 8, the flame front also moves slightly upstream into the mixing tube and tends to interact with the fuel injection system and form a small recirculation bubble ahead of flame front (like 30%$H_2$ ) at t=13.4s (Fig. 8). Since the further upstream movement of the flame front is annihilated due to unfavorable condition, the flame front is actually pushed back by the incoming flow (t=13.8s & t=17.6s in Fig. 8) and the small recirculation bubble (t=13.4s in Fig. 8) also disappears at t=13.8s & t=17.6s. This back-n-forth movement of the flame front creates a quasi-stable (periodic) situation without occurrence of flashback for $CH_4$ and produces the limit cycle behavior in the pressure fluctuations observed in Fig. 5(b).

Fig. 9(a) shows the axial velocity fluctuations at further downstream of dump plane. The velocity fluctuations in Fig. 9(a) also produce same frequency (Fig. 5(a)) but disappears at later time instants as observed in the axial velocity spectra in Fig. 9(b) where Strouhal number is calculated as $St=fD/U_o$ (D=center-body diameter, $U_o$=inlet bulk velocity). In addition, the velocity fluctuations (Fig. 9(a)) show some wavy pattern with a time period of ~1s. The corresponding frequency appears to be the order of 1 Hz ($St$~0.0017) This is coming due to the interaction of flame front with counter rotating vortex pairs that wrinkle the flame front downstream. The counter rotating vortex pairs are the vortex sheet on both sides of the bluff body, which roll up into regions of intense vorticity (Fig. 10). This wrinkling and corrugation of the flame sheet is caused due to unstable, separated shear layer of the bluff body, also partly due to the flame position which nearly lies parallel to the flow field. As flame generated vorticity (FGV) counterbalances rotating vortex pairs, the wrinkling of flame front is also attenuated and flame front detaches while the whole process repeats again. The periodicity of this wrinkling process shows up in Fig. 9(a) (~1s).

The effects of heat release on the flow field can be explained using the vorticity transport equation (Eq. 26) which essentially shows the evolution of the vorticity of a moving fluid element in space and is written as:

$$\frac{D\vec{\omega}}{Dt} = \underbrace{(\vec{\omega}.\vec{\nabla})\vec{u}}_{I} - \underbrace{\vec{\omega}(\vec{\nabla}.\vec{u})}_{II} - \underbrace{\frac{\vec{\nabla}p \times \vec{\nabla}\rho}{\rho^2}}_{III} + \underbrace{\nu\nabla^2\vec{\omega}}_{IV} \tag{27}$$

where the RHS terms are: (I) Vortex stretching, (II) Gas expansion, (III) Baroclinic production, and (IV) Viscous diffusion. The terms (I) and (IV) have influence regardless of reacting or non-reacting flows. The term (IV) sharply rises across the flame due to change in temperature, and thus enhances the rate of diffusion and dampens the vorticity. However, the misalignment of the pressure and density gradients due to inclination and expansion of the flame with respect to the flow field contributes to the baroclinic production (III) of vorticity as shown in Fig. 11. The gas expansion term (II), acts as a sink in reacting cases, is directly proportional to the gas dilatation ratio across the flame ($\rho_u/\rho_b$) and increases as the temperature increases in presence of combustion. Hence these two terms (II & IV) stabilize each other influences in reacting flow field (Fig. 12). Therefore, the production of negative aximuthal vorticity (Fig. 12) at the inner edge of the flame (burnt region along the pipe wall) and also along the flame surface, is primarily due to interaction between shear generated vorticity and flame generated, baroclinic vorticity.

Moreover, it is worthwhile to mention that the negative azimuthal vorticity induces a negative axial velocity in the flow filed following the relationship as given

$$w_{ind}(x) = \frac{1}{2} \int\limits_{-\infty}^{\infty} \int\limits_{0}^{\infty} \frac{r'^2 \, \omega_\theta(r',x')}{\left[ r'^2 + (x-x')^2 \right]^{3/2}} dr' dx' \tag{28}$$

Hence this clearly states that the greater the vorticity, the greater is the induced velocity. Thus, this induced negative velocity pushes the stagnation point ahead of the flame tip (Fig. 7, t=8s–t=10.5s) further upstream and helps to form a small recirculation bubble primarily due to pressure jump across the convex flame orientation in the flow field (Fig. 11). The formation of this small recirculation bubble ahead of the flame tip gives rise to the generation of greater negative azimuthal vorticity, which in turn, produces higher induced negative velocity and the flame tip movement to further upstream becomes completely uncontrollable. That's why 30%$H_2$ exhibits severe upstream movement while $CH_4$ does not show the same behavior (Figs. 7 & 8). In order to further illustrate the pressure jump, which causes to form recirculation bubble ahead of flame tip and giving rise to the negative induced velocity, across the flame front in the axial and radial direction (Fig. 11), the simplified radial momentum equation is used: $\dfrac{\partial p}{\partial r} = \dfrac{\rho \overline{w}^2}{r}$, where $\rho$ is the density, $\overline{w}$ is the tangential velocity, and r is the radial coordinate [18]. Assuming p=p(x,r), radial pressure gradient can be related as: $\left. \dfrac{dp}{dr} \right|_{r_i}^{r_o} = \int\limits_{r_i}^{r_o} \dfrac{\rho \overline{w}^2}{r} dr$, where $r_i$ is the radius at the center body wall and $r_o$ is the radius at the outer wall. Due to density difference between outer and inner wall (Fig. 11) across the flame front in the mixing tube, the radial pressure difference is always dp>0. To maintain conservation of momentum, the axial pressure gradient changes. However, the axial pressure gradient along the inner wall is higher than that along the outer wall as shown below:

$$\frac{dp}{dx}\bigg|_{r_o} - \frac{dp}{dx}\bigg|_{r_i} = \int_{r_i}^{r_o} \frac{\partial}{\partial x}\left(\frac{\rho \overline{w}^2}{r}\right) dr \Rightarrow \quad \frac{dp}{dx}\bigg|_{r_i} = \frac{dp}{dx}\bigg|_{r_o} - \int_{r_i}^{r_o}\left(\overline{w}^2 \frac{\partial \rho}{\partial x} + 2\rho \overline{w} \frac{\partial \overline{w}}{\partial x}\right) dr \quad (29)$$

The above expression clearly reveals that the axial pressure gradient along the inner wall is larger than that at outer wall due to decrease in density in the axial direction ($\frac{\partial \rho}{\partial x} < 0$, Fig. 11) and decrease of tangential velocity in the axial direction as well. Therefore, as observed in Fig. 7, the positive pressure gradient along the inner wall pushes the flame tip into the mixing tube and starts interacting with incoming cross flow fuel injection (Fig. 7, t=8.0s-t=10.5s) resulting formation of low velocity recirculation region ahead of flame tip which in turn gives rise to the production of negative axial velocity to favor this upstream propagation severely. Whereas in case of $CH_4$, the inner wall pressure gradient becomes negative and pushes the flame front back at later time instants (Fig. 8; t=13.8s, 17.6s), never leads to flashback.

A more detailed observation of vorticity budget terms (Eq. 27) supports the above phenomena, which contributes to the generation of negative azimuthal vorticity, in turn produces the induced negative velocity. Figure 12 shows the distribution of change in budget terms between t=8s to t=12.0s for 30%$H_2$ mixture. The thorough analysis clearly exhibits that the combined effects of vortex stretching and baroclinic production give rise to the negative azimuthal vorticity particularly along the flame front, while the vortex expansion and diffusion terms stabilize each other influences. Since negative azimuthal vorticity induces negative axial velocity, vortex stretching and baroclinic production are primarily responsible for upstream flame propagation and leading to flashback. Whereas for $CH_4$, baroclinic production and vortex stretching contribute to the production of positive azimuthal vorticity along the flame front and that's why flame front is pushed backed as shown in Fig. 8; never leads to flashback and produces quasi stable back-n-forth movement of flame front for this case.

However, it is noteworthy to mention that the previous studies [12-16, 18-19] on different swirl stabilized burner reported CIVB driven flashback where the time scale for flashback was in the order of milliseconds for a laboratory scale burners fired with $CH_4$. In contrast, flame propagation velocity in the present case turns out to be ~0.007m/s which is very less compared to the previous studies in this area, though it has been found to be in order of few centimeters/s for higher flow rates (higher thermal power). The primary difference arises due to the burner configuration between the earlier studies [18-19] and the present study. In that burner [18-19], fuel & air has been premixed at a section which is far away from dump plane while in the present case we have a jet in cross flow injection close to the dump plane and that has lot of impact. That's why the flow dynamics has been presented first and then followed by the discussion on upstream flame movement. This is slightly misleading to find out a direct correlation between these two burners. Because, once the burner configuration is different, the flame propagation speed along with the physical effect would definitely differ. As noted earlier, cross flow fuel injection initially prohibiting the flame to move upstream: (a) due to lack of pre-mixedness at the beginning, and (b) low velocity region created along the center-body right after injection point along with the influence of sudden expansion of swirled air. In combination of these interactions make the system is very complicated while, in comparison, the other burner configuration [18-19] appears to be simple. That's why we have observed different time scale for upstream flame movement.

In summary, the basic idea of this CIVB is that the flame contributes to vortex breakdown, and results a low or negative flow region ahead of it (recirculation bubble formation with a stagnation point). As the flame tip moves forward, causing the location of vortex breakdown region to advance further upstream. This process continues as the flame proceeds further and further upstream. It has been also reported that in CIVB, flashback can occur even if turbulent flame speed ($S_T$) is everywhere less than the flow velocity [59].

## 5. CONCLUSIONS

LES with a TF model is used to investigate hydrogen enriched premixed flames in a laboratory based model combustor. The effect of hydrogen addition is mainly studied, especially in the context upstream flame propagation dynamics. Upstream flame propagation driven by Combustion Induced Vortex Breakdown (CIVB) is known to be a severe problem for premixed swirl burners, and has been investigated in great details. A 2-step chemical scheme for methane combustion and 1-step for hydrogen combustion are invoked to represent the flame chemistry for methane-hydrogen-air combustion. The equivalence ratio for the flame is 0.7 and the geometric swirl number for the configuration is 0.82. This study leads to the following conclusions:

1. LES-TF model is able to properly capture hydrogen enriched combustion behavior. Both the reacting and non-reacting velocity profiles are well predicted.

2. Hydrogen enrichment reduces the size and shape of the recirculation bubble.

3. Hydrogen enrichment also leads to flame flashback due to higher combustibility. Combustion-Induced-Vortex-Breakdown (CIVB) is the main mechanism behind the upstream flame propagation and is governed by the interaction of turbulence and the chemistry due to its influence on the vorticity.

4. The gaseous combustion and associated temperature influences the gas-expansion and baroclinic production. The vortex stretching and baroclinic production mechanisms contribute to the net vorticity generation which produces considerable levels of negative axial velocity that favors upstream flame propagation.

5. For low hydrogen content flames, the flame does not produce enough baroclinic torque, and the combined effects of gas-expansion and diffusion can stabilize the vortex flow and prevent upstream flame movement. This is what happens in the case of pure $CH_4$ flame, and flashback is not observed.

This study demonstrates that the Thickened-Flame based LES approach with simplified chemistry for reacting flows is a promising tool to investigate reacting flows in complex geometries.

## ACKNOWLEDGMENTS

The authors would like to thank Shengrong Zhu for providing the experimental data. This work was supported by the Clean Power and Energy Research Consortium (CPERC) of Louisiana through a grant from the Louisiana Board of Regents. Simulations were carried out on the computers provided by LONI network at Louisiana, USA (www.loni.org) and HPC resources at LSU, USA (www.hpc.lsu.edu). Finally, the manuscript preparation as well as partial analysis of data is carried out using the resources available at Indian Institute of Technology Kanpur (IITK), India. This support is gratefully acknowledged.

## Nomenclature

A        pre-exponential constant
$C_s$        LES model coefficient
$D_i$        molecular diffusivity
E        efficiency function

| | |
|---|---|
| $E_a$ | activation energy |
| $S_{ij}$ | mean strain rate tensor |
| $T_a$ | activation temperature |
| U | mean axial velocity |
| $U_o$ | bulk inlet velocity |
| $u_i$ | velocity vector |
| $u'$ | rms turbulence velocity |
| W | mean tangential velocity |
| w' | tangential RMS velocity |
| $x_i$ | Cartesian coordinate vector |
| $Y_i$ | species mass fraction |

**Greek symbols**

| | |
|---|---|
| $\Delta$ | mesh spacing |
| $\nu_t$ | kinematic turbulent eddy viscosity |
| $\bar{\rho}$ | mean density |
| $\omega_i$ | reaction rate |

# REFERENCES

1. Calvetti S, Carrai L, Cecchini D. Progettazione di un Combustore DLN Prototipoper TG per la Co-Combustione di Gas Naturale e Syngas da Biomassa. Technical report, Enel Produzione, Pisa, Italy, ENELP/RIC/RT/-2001/146/-IT+RT.RIC.PI; 2001.

2. Griebel P, Boscheck E, Jansohn P. Lean Blowout Limits and Nox Emissions of Turbulent Lean Premixed Hydrogen-Enriched Methane/Air Flames at High Pressure. J. Eng. Gas Turbines Power 2007; 129: 404-10.

3. Mariotti M, Tanzini G, Faleni M, Castellano L. Sperimentazione di Fiamme di Idrogeno a Pressione Atmosferica in un Combustore per Turbogas con Iniezione di Inerti. Technical report, Enel Produzione, Pisa, Italy, ENELP/RIC/RT/-2202/0063; 2002.

4. Wohl K. Quenching, Flash-Back, Blow-off Theory and Experiment. $4^{Th}$ Symp. (Int.) Combust. 1952; 69-9.

5. Plee SL, Mellor AM. Review of Flashback Reported in Prevaporizing/Premixing Combustor. Combust. Flame 1978; 32: 193-03.

6. Putnam A, Jensen R. Application of Dimensionless Numbers to Flashback and other Combustion Phenomena. Proc. Combust. Inst. 1948; 125: 89-8.

7. Noble DR, Zhang Q, Lieuwen T. Hydrogen Effects Upon Flashback and Blowout. Proc. ICEPAG 2006; ICEPAG2006-24102.

8. Fritz J, Kröner M, Sattelmayer T. Flashback in a swirl Burner with Cylindrical Premixing Zone. J. Eng. Gas Turbines Power 2004; 126: 276-83.

9. Kröner M, Fritz J, Sattelmayer T. Flashback Limits for Combustion Induced Vortex Breakdown in a swirl Burner. J. Eng. Gas Turbines Power 2003; 125: 693-00.

10. Kiesewtter F, Hirsch C, Fritz M, Kröner M, Satterlmayer T. Two-Dimensional Flashback Simulation in Strongly Swirling Flows. ASME Turbo Expo 2003; GT2003-38395.

11. Sommerer Y, Galley D, Poinsot T, Ducruix S, Lacas F, Veynante D. Large eddy simulation and experimental study of flashback and blow-off in a lean partially premixed swirled burner. J. Turbulence 2004; 5(37): 61-8.

12. Nauert A, Pettersson P, Linne M, Dreizler A. Experimental analysis of flashback in lean premixed swirling flames: conditions close to flashback. Exp. Fluids 2007; 43: 89-00.

13. Kiesewtter F, Knole M, Satterlmayer T. Analysis of Combustion Induced Vortex Breakdown Driven Flame Flahback in a Premix Burner with Cylindrical Mixing one. J. Eng. Gas Turbines Power 2007; 129: 929-36.

14. Kröner M, Satterlmayer T, Fritz M, Kiesewtter F, Hirsch C. Flame Propagation in Swirling Flws-Effect of Local Extinction on the Combustion Induced Vortex Breakdown. Combust. Sci. Technol. 2007;179: 1385-16.

15. Knole M, Kiesewtter F, Satterlmayer T. Simultaneous high repetition rate PIV-LIF measurements of CIVB driven flashback. Exp. Fluids 2008; 44: 529-38.

16. Knole M, Satterlmayer T. Time Scale Model for the prediction of the onset of flame flashback driven by combustion induced vortex breakdown (CIVB). ASME Turbo Expo 2009; GT2009-59606.

17. Voigt T, Habusreuther P, Zarzalis N. Simulation of Vorticity Driven Flame Instability Using A Flame Surface Density Approach Including Markstein Number Effects. ASME Turbo Expo 2009; GT2009-59331.

18. Heeger C, Gordon RL, Tummers M, Sattelmayer T, Dreizler A. Experimental analysis of flashback in lean premixed swirling flames: upstream flame propagation. Exp. Fluids 2010; 49: 853-63.

19. Tangermann E, Pfitzer M, Knole M, Sattelmayer T. Large-Eddy Simulation and experimental observation of Combustion-Induced Vortex Breakdown. Combust. Sci. Technol. 2010; 182: 505-16.

20. Morris JD, Symonds RA, Ballard FL, Banti A. Combustion aspects of application hydrogen natural gas fuel mixtures to MS9001EDLN-1 gas turbines at Elsta plant, Terneuzen, The Netherlands. ASME Turbo Expo 2009; GT1998-359.

21. Schefer RW. Hydrogen enrichment for Improved Lean Flame Stability. Int. J. Hydrogen Energy 2003; 28(10): 1131-41.

22. Schefer RW, Wicksall DM, Agrawal AK. Combustion of Hydrogen-Enriched Methane in a Lean Premixed Swirl-Stabilized Burner. Proc. Combust. Inst. 2002; 29: 843-51.

23. Kim HS, Arghode VK, Linck MB, Gupta AK. Hydrogen addition effects in a confined swirl-stabilized methane-air flame. Int. J. Hydrogen Energy 2009; 34: 1054-62.

24. Kim HS, Arghode VK, Gupta AK. Flame Characteristics of hydrogen-enriched methane-air premixed swirling flames. Int. J. Hydrogen Energy 2009; 34: 1062-73.

25. Ballester J, Hernández R, Sanz A, Smolarz A, Barroso J, Pina, A. Final Report on the Development of a Hydrogen-Fueled Combustion Turbine Cycle for Power Generation. J. Engg. Gas Turbines Power 2009; 121: 38-5.

26. Strakey P, Sidwell T, Ontko J. Investigation of the effects of hydrogen addition on lean extinction in a swirl stabilized combustor. Proc. Combust. Inst. 2007; 31: 3173-80.

27. Tuncer O, Acharya S, Uhm JH. Dynamics, Nox and flashback characteristics of confined premixed hydrogen-enriched methane flames. Int. J. Hydrogen Energy 2009; 34: 496-06.

28. Schneider C, Dreizler A, Janicka J. Fluid Dynamical Analysis of Atmospheric Reacting and Isothermal Swirling Flows. Flow Turbulence Combust. 2005; 74: 103-27.

29. Syred DG, Beer JM. Combustion in swirling flows: A review. Combust. Flame 1974; 23: 143-01.

30. De A, Zhu S, Acharya S. An experimental and computational study of a swirl stabilized premixed flame. ASME Turbo Expo 2009; GT2009-60230.

31. Syred N. A review of oscillation mechanisms and the role of preceesing vortex core (PVC) in swirl combustion systems. Prog. Energy Combust. Sci. 2006; 32: 93-61.

32. Duwig C, Fuchs L, Lacarelle A, Beutke M, Paschereit CO. Study of the vortex breakdown in a conical swirler using LDV, LES and POD. ASME Turbo Expo 2007; GT2007-27006.

33. Hawkes E, Cant R. Implications of a flame surface density approach to large eddy simulation of premixed turbulent combustion. Combust. Flame 2001; 126: 1617-29.

34. Williams FA. Combustion Theory. Benjamin/Cummins; Menlo park, CA: 1985.

35. Peters N. Tubulent Combustion. Cambridge Univ. Press; London/New York: 2000.

36. Düsing M, Sadiki A, Janicka J. Towards a classification of models for the numerical simulation of premixed combustion based on a generalized regime diagram. Combust. Theory Model. 2006; 10: 105-32.

37. Pitsch H, Duchamp de La Geneste L. Large-eddy simulation of a premixed turbulent combustion using level-set approach. Proc. Combust. Inst. 2002; 29: 2001-08.

38. Pope SB. Pdf methods for turbulent reactive flows. Prog. Energy Combust. Sci. 1985; 11:119-92.

39. Colin O, Ducros F, Veynante D, Poinsot T. A thickened flame model for large eddy simulation of turbulent premixed combustion. Phys. Fluids 2000; 12(7): 1843-63.

40. Germano M, Piomelli U, Moin, P, Cabot WH. A dynamic subgrid-scale eddy viscosity model. Phys. Fluids 1991; 3: 1760-65.

41. Smagorinsky J. General circulation experiments with the primitive equations. I: The basic experiment. Month. Weather Rev. 1963; 91: 99-64.

42. Poinsot T, Veynante D. Theoretical and Numerical Combustion. Edwards: 2001.

43. Künne G, Ketelheun A, Janicka J. LES modeling of premixed combustion using a thickened flame approach coupled with FGM tabulated chemistry. Combust. Flame 2011; 158 (9): 1750-67.

44. Butler TD, O'Rourke PJ. A numerical method for two-dimensional unsteady reacting flows. Proc. Combust. Inst. 1977; 16: 1503-15.

45. Kuo, Kenneth K. Principles of Combustion, Second edition. John Wiley & Sons. Inc.: 2005.

46. Charlette F, Meneveau C, Veynante, D. A Power-Law flame wrinkling model for LES of premixed turbulent combustion, Part I: Non Dynamic formulation and initial tests. Combust. Flame 2002; 131: 159-80.

47. Charlette F, Meneveau C, Veynante D. A Power-Law flame wrinkling model for LES of premixed turbulent combustion, Part II: Dynamic formulation and initial tests. Combust. Flame 2002; 131: 181-97.

48. Durand L, Polifke W. Implementation of the thickened flame model for large eddy simulation of turbulent premixed combustion in a commercial solver. ASME Turbo Expo 2007; GT2007-28188.

49. De A, Acharya S. Large Eddy Simulation of Premixed Combustion with a Thickened-Flame Approach. ASME Turbo Expo 2008; GT2008-51320.

50. De A, Acharya S. Large Eddy Simulation of Premixed Bunsen flame using a modified Thickened-Flame model at two Reynolds number. Combust. Sci. Technol. 2009; 181(10): 1231-72.

51. Selle L, Lartigue G, Poinsot T, Koch R, Schildmacher KU, Krebs W, et al. Compressible large eddy simulation of turbulent combustion in complex geometry on unstructured meshes. Combust. Flame 2004; 137: 489-05.

52. DOE Final Report. Premixer Design for High Hydrogen Fuels. DOE Cooperative Agreement No. DE-FC26-03NT41893.DE-FC26-03NT41893: 2005.

53. Kee RJ, Rupley FM, Miller JA. Chemkin-II: A Fortran Chemical Kinetics Package for Analysis of Gas-Phase Chemical Kinetics. Sandia Report; SAND 89-8009B.

54. Kee RJ, Dixon-Lewis, Warnats J, Coltrin ME, Miller JA. A Fortran Computer code for the evaluation of Gas-Phase multicomponent transport properties. Sandia Technical Report; SAND 86-8246.

55. Weiss JM, Smith WA. Preconditioning applied to variable and constant density flows. AIAA J. 1995; 33: 2050-57.

56. Edwards JR. A low-diffusion flux-splitting scheme for Navier-Stokes calculations. Comput. Fluids 1997; 26: 635-59.

57. Akselvoll K, Moin P. Large-eddy simulation of turbulent confined coannular jets. J. Fluid Mech. 1996; 315: 387-11.

58. Galpin J, Naudin A, Vervisch L, Angelberger C, Colin O, Domingo P. Large-eddy simulation of a fuel-lean premixed turbulent swirl burner. Combust. Flame 2008; 155: 247-66.

59. Lieuwen T, MacDonell V, Peterson E, Santavicca D. Fuel flexibility influences on premixed combustor Blowout, Flahsback, Autoignition, and stability. J. Eng. Gas Turbines Power 2008; 130(1): 011506.

**List of Figure captions**

Figure 1. Schematic view of computational domain for the swirl injector

Figure 2. Non-reacting flow results for Re=10144 at different axial locations [D=Center-body diameter]: Experimental data ($\triangle$), Lines are LES predictions: fine mesh (—), coarse mesh ($\cdots$). Mean axial velocity $U/U_o$ , Mean tangential velocity $W/U_o$ , Axial velocity fluctuation $u_{rms}/U_o$ , Tangential velocity fluctuation $w_{rms}/U_o$.

Figure 3. Reacting flow results for Re=10144: (a) Swirl number (top) and recirculation bubble size (bottom) at different $H_2$% [X1, X2, X3, X4=(X/2D): 0.40, 0.79, 1.18, 1.58], (b) Radial velocity profiles with 30%$H_2$ at different axial locations: Experimental data ($\triangle$), Lines are LES predictions: Mean axial velocity $U/U_o$ , Axial velocity fluctuation $u_{rms}/U_o$

Figure 4. (a) Time averaged temperature contours (K scale) at different $H_2$%, (b) Instantaneous flame propagation for $CH_4$ & 30%$H_2$ (Temperature in K scale)

Figure 5. (a) Frequency spectra at the upstream of the dump plane, (b) Pressure fluctuations at the upstream of the dump plane

Figure 6. Instantaneous flame movement at phases 1-5 of oscillation frequency (45Hz) before flashback with 30%$H_2$: stream lines colored with temperature contours (K scale)

Figure 7. Instantaneous flame movement during flashback with 30%$H_2$: Stream lines colored with temperature contours (K scale), isotherms (black lines: 300K, 675K, 1050K, 1425K)

Figure 8. Instantaneous flame movement with $CH_4$: Stream lines colored with temperature contours (K scale), isotherms (black lines: 300K, 675K, 1050K, 1425K)

Figure 9. (a) Axial velocity fluctuations at further downstream (wrinkled flame location) of dump plane, (b) Axial velocity spectra (12.5-17.5s) at further downstream (wrinkled flame location) of dump plane

Figure 10. Flame-Vortex interaction at downstream of dump plane: Time between frames (0.2s), Red is positive vorticity (50 1/s), Blue is negative vorticity (-50 1/s), Isotherm level at T=1500K

Figure 11. Pressure ($p_o$=101325 Pa), density and baroclinic production plots during flashback with 30%$H_2$

Figure 12. Individual terms of vorticity transport equation (Eq. 27) along the flame arc length of isotherm (1050K) during flashback with 30%$H_2$ : 0=center-body wall (flame tip)

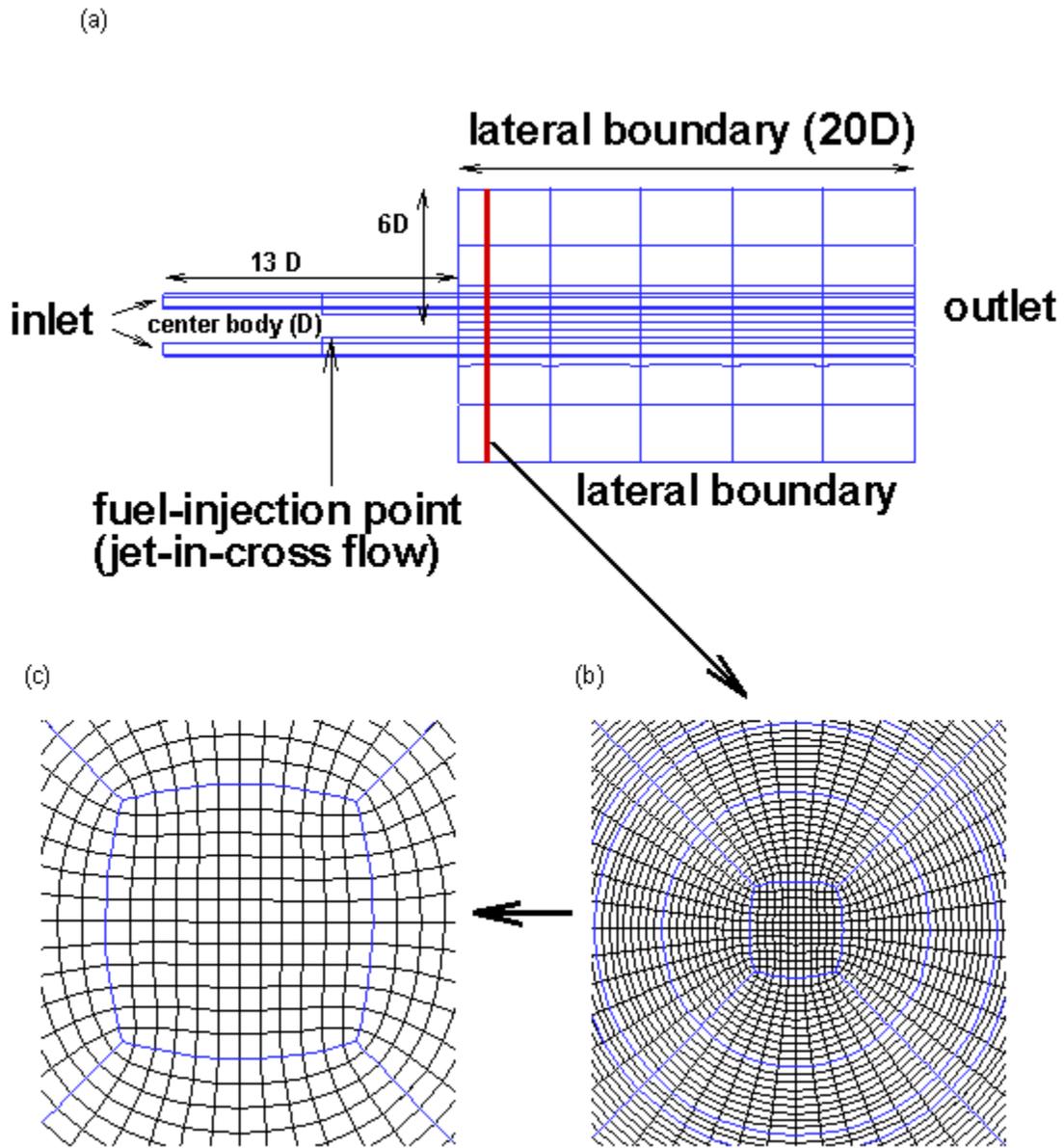

Figure 1

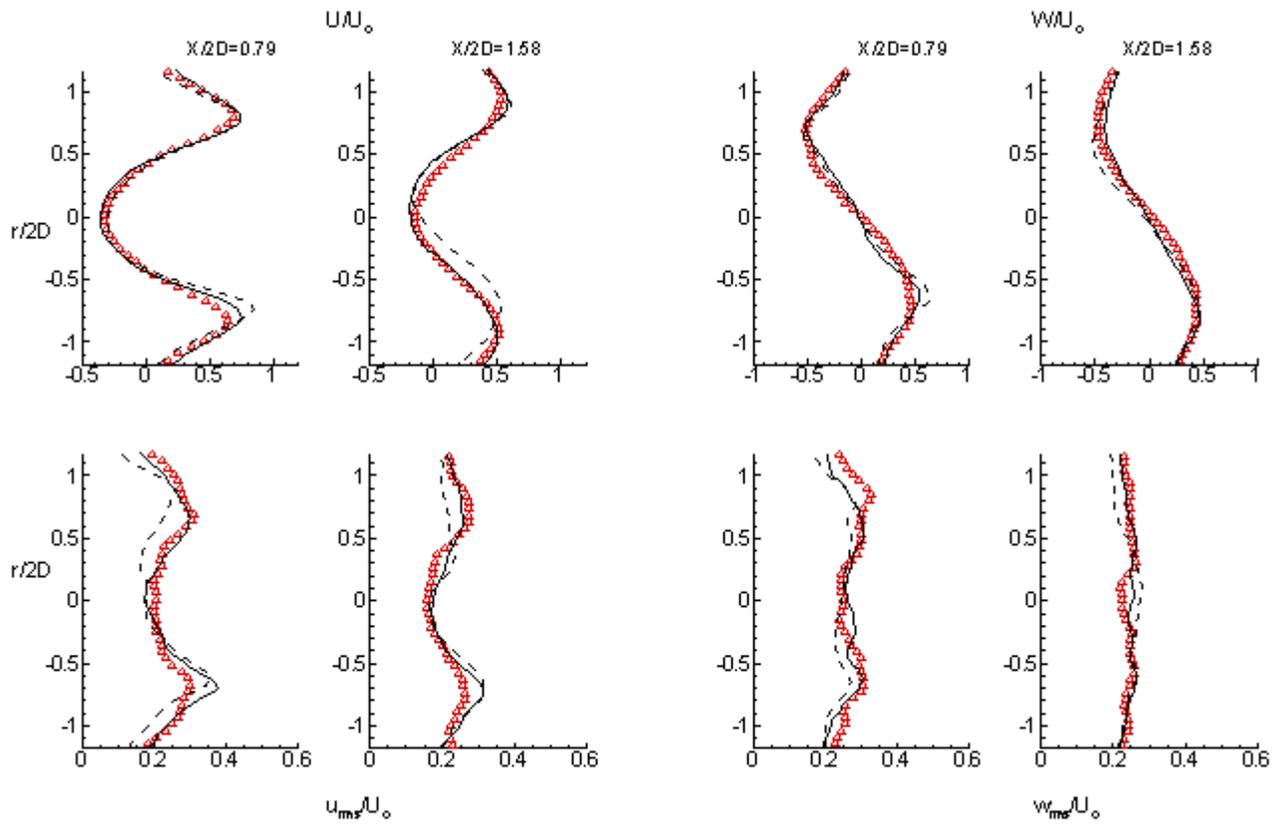

Figure 2

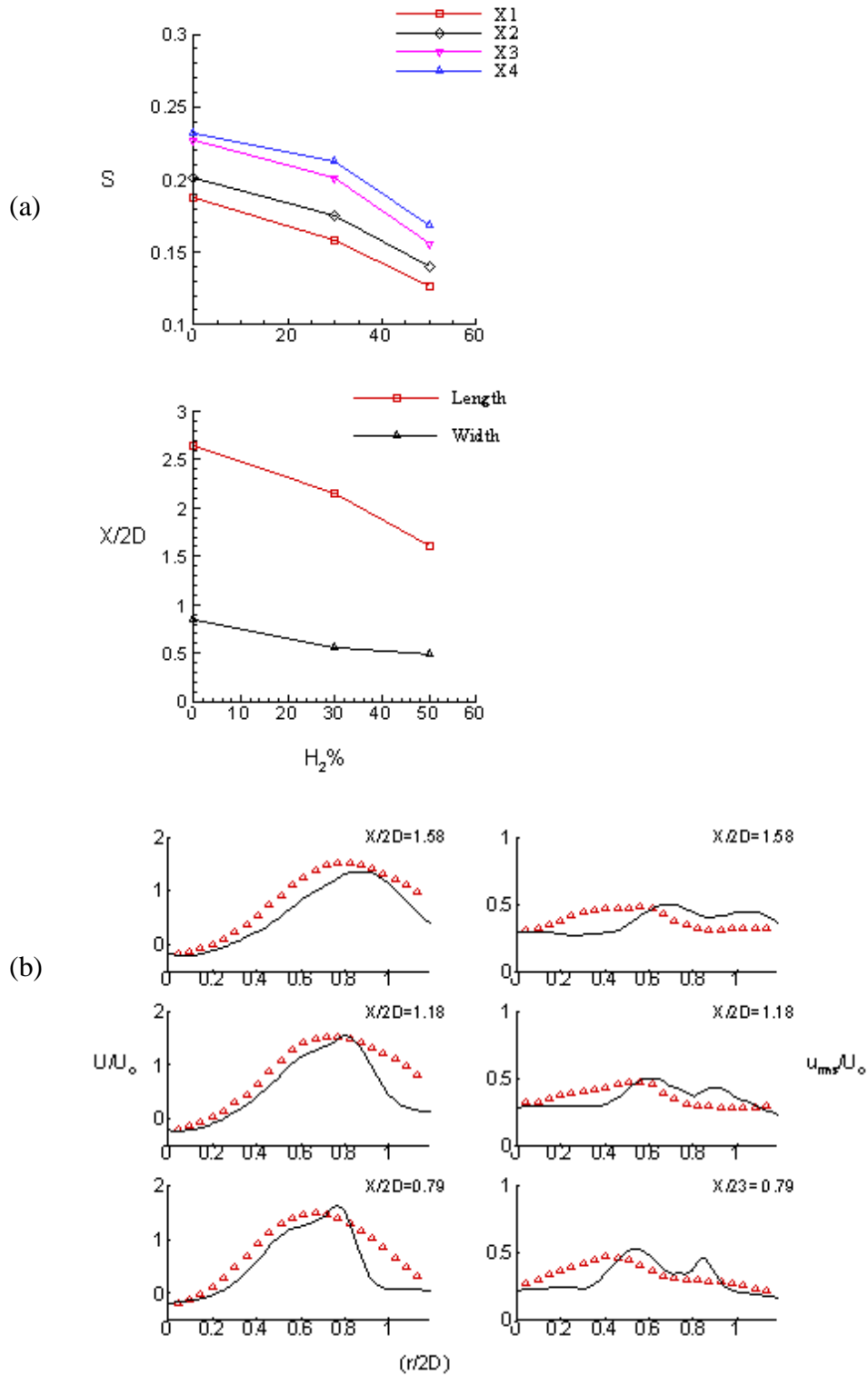

(a)

(b)

Figure 3

(a)

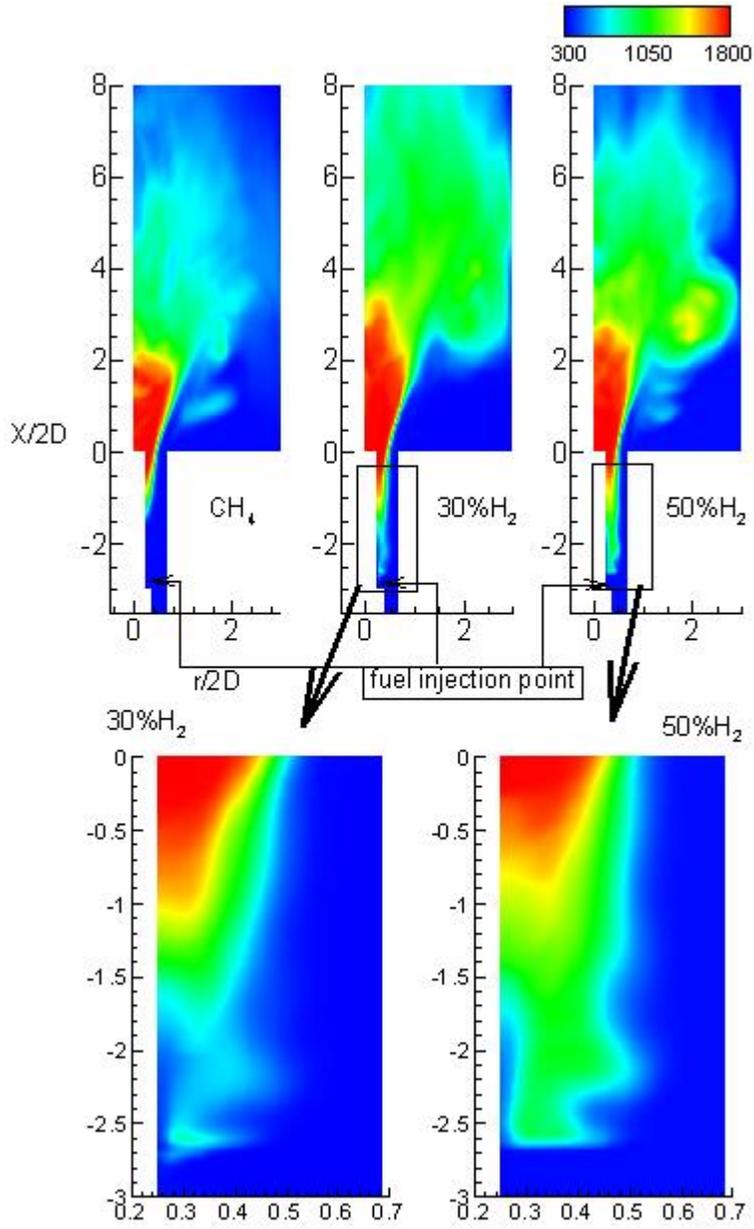

(b)

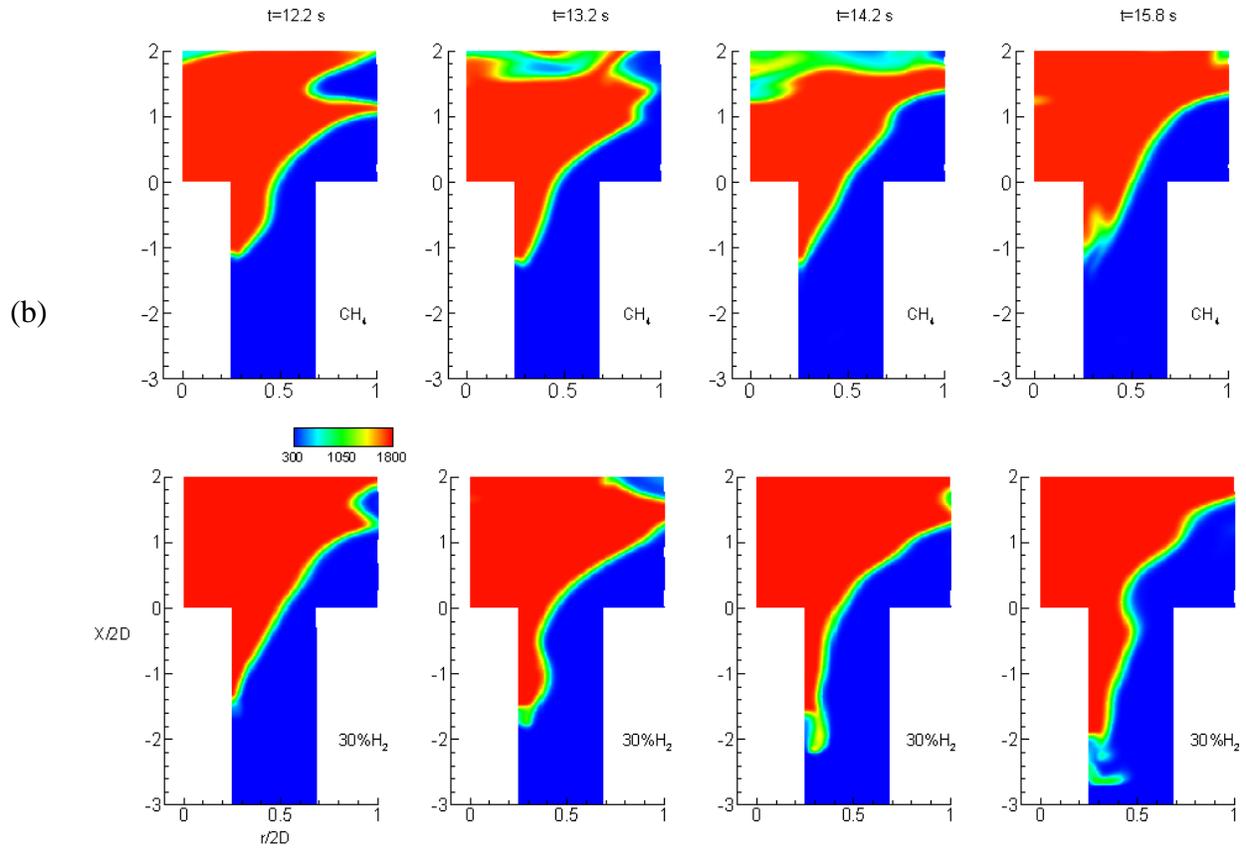

Figure 4

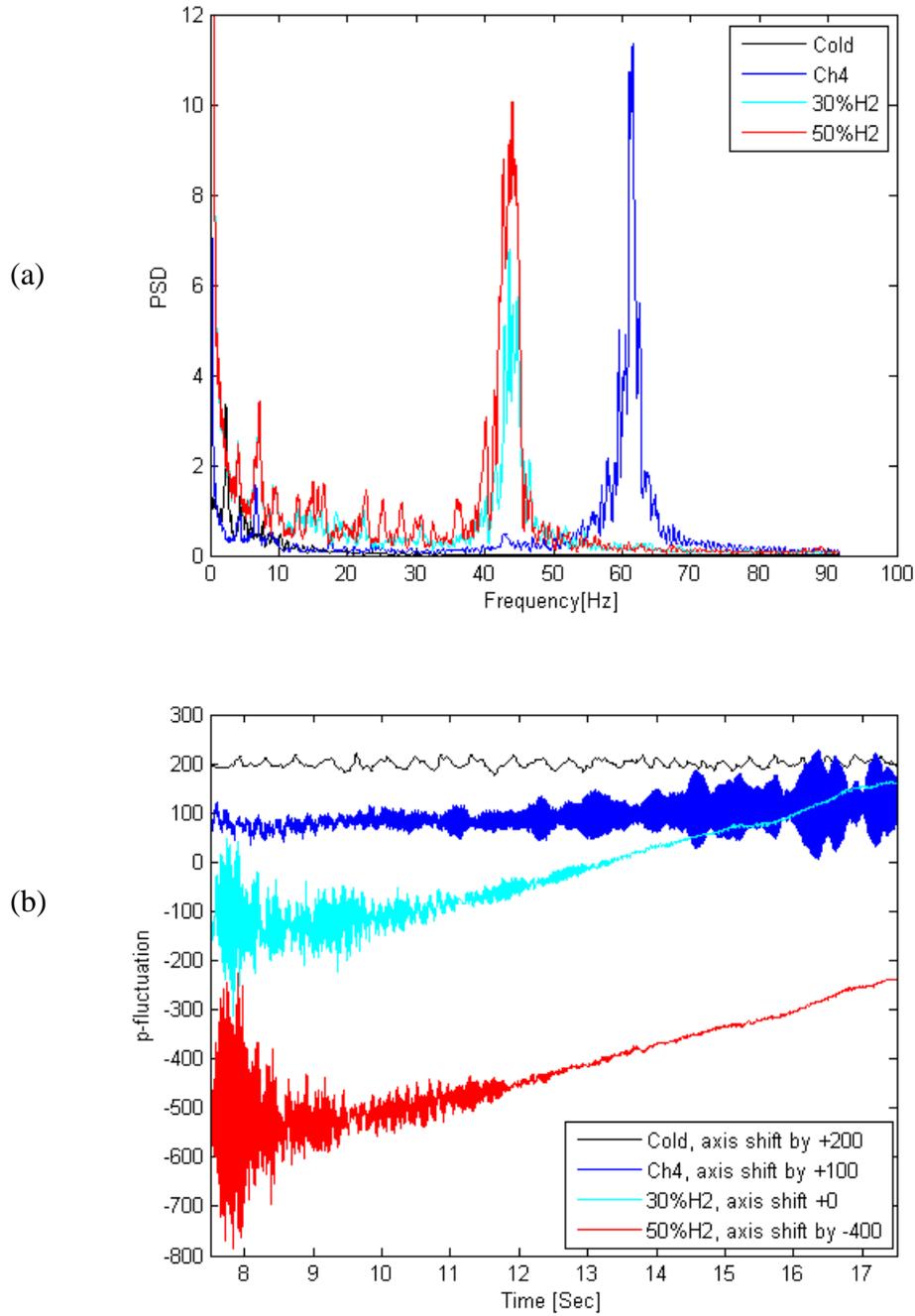

(a)

(b)

Figure 5

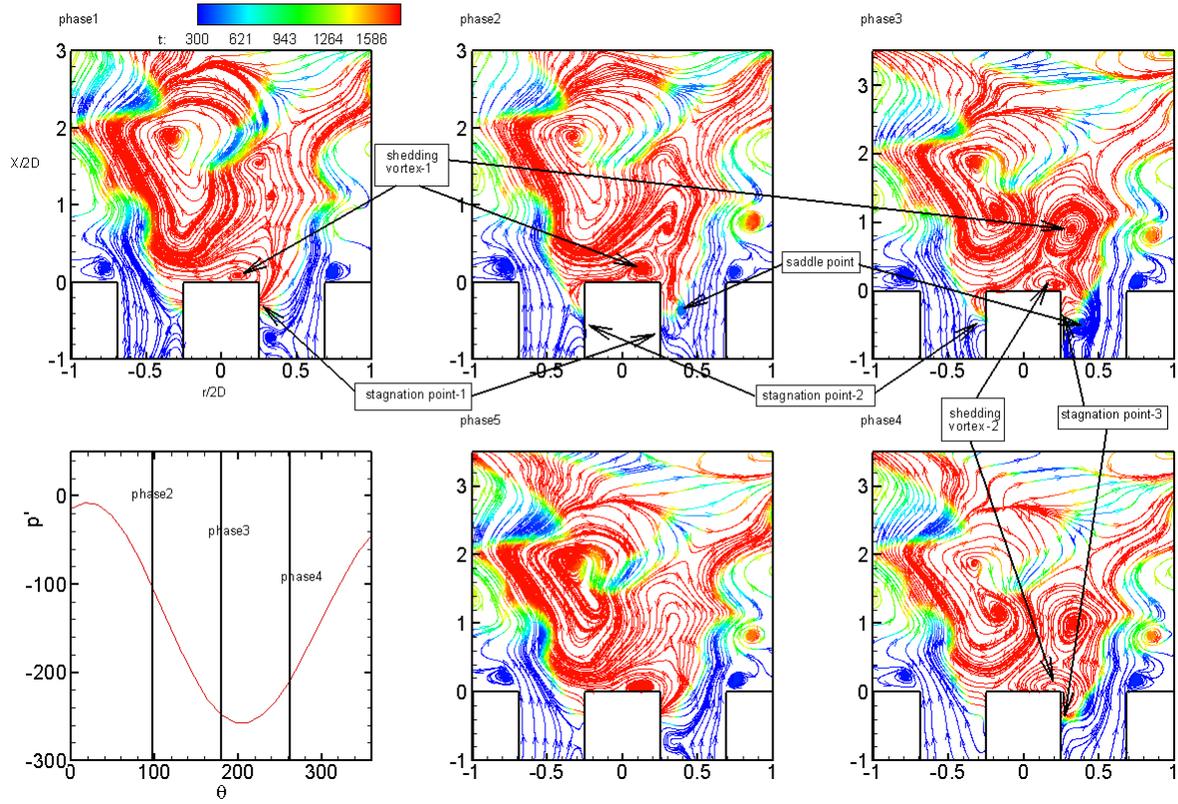

Figure 6

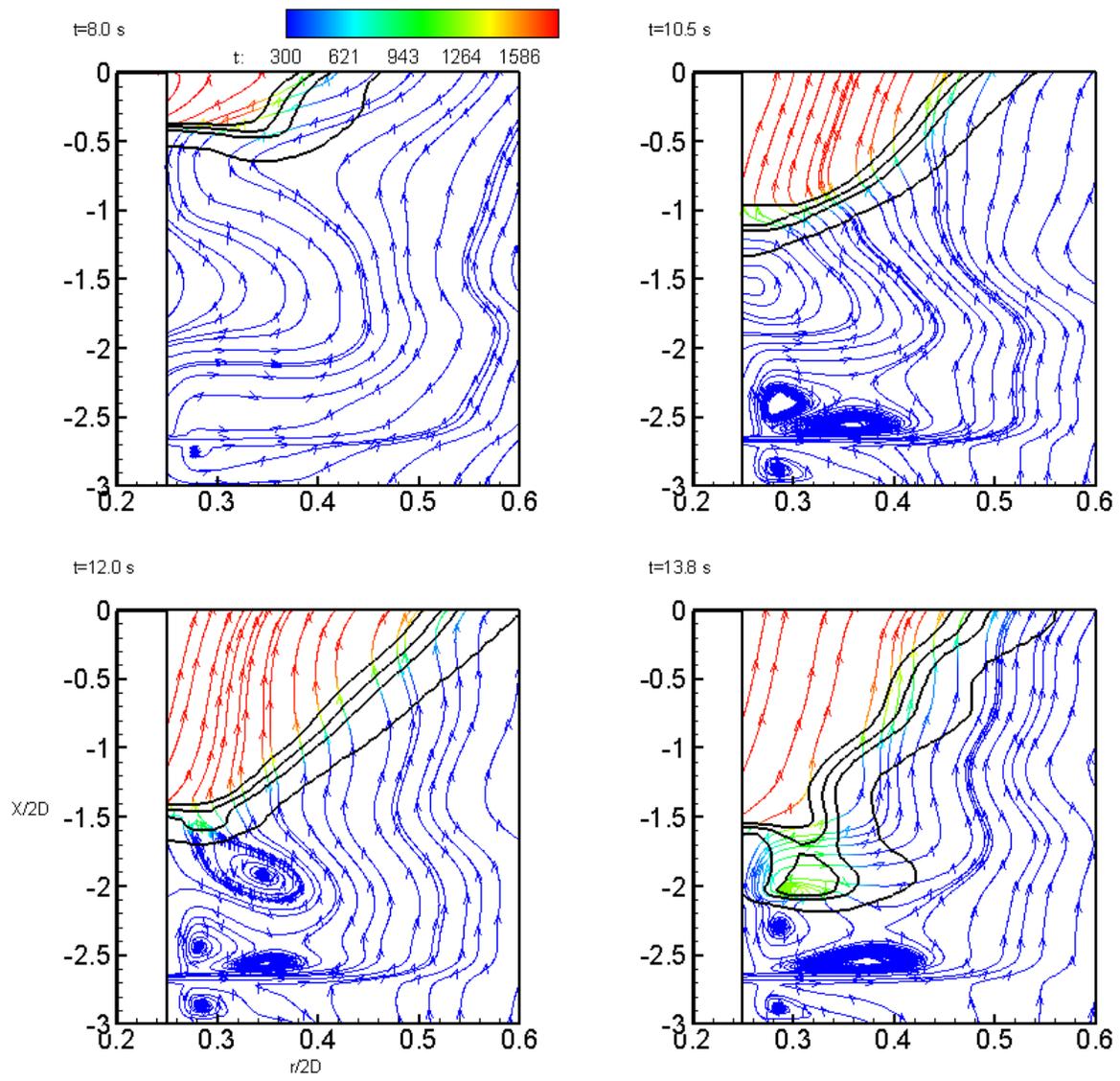

Figure 7

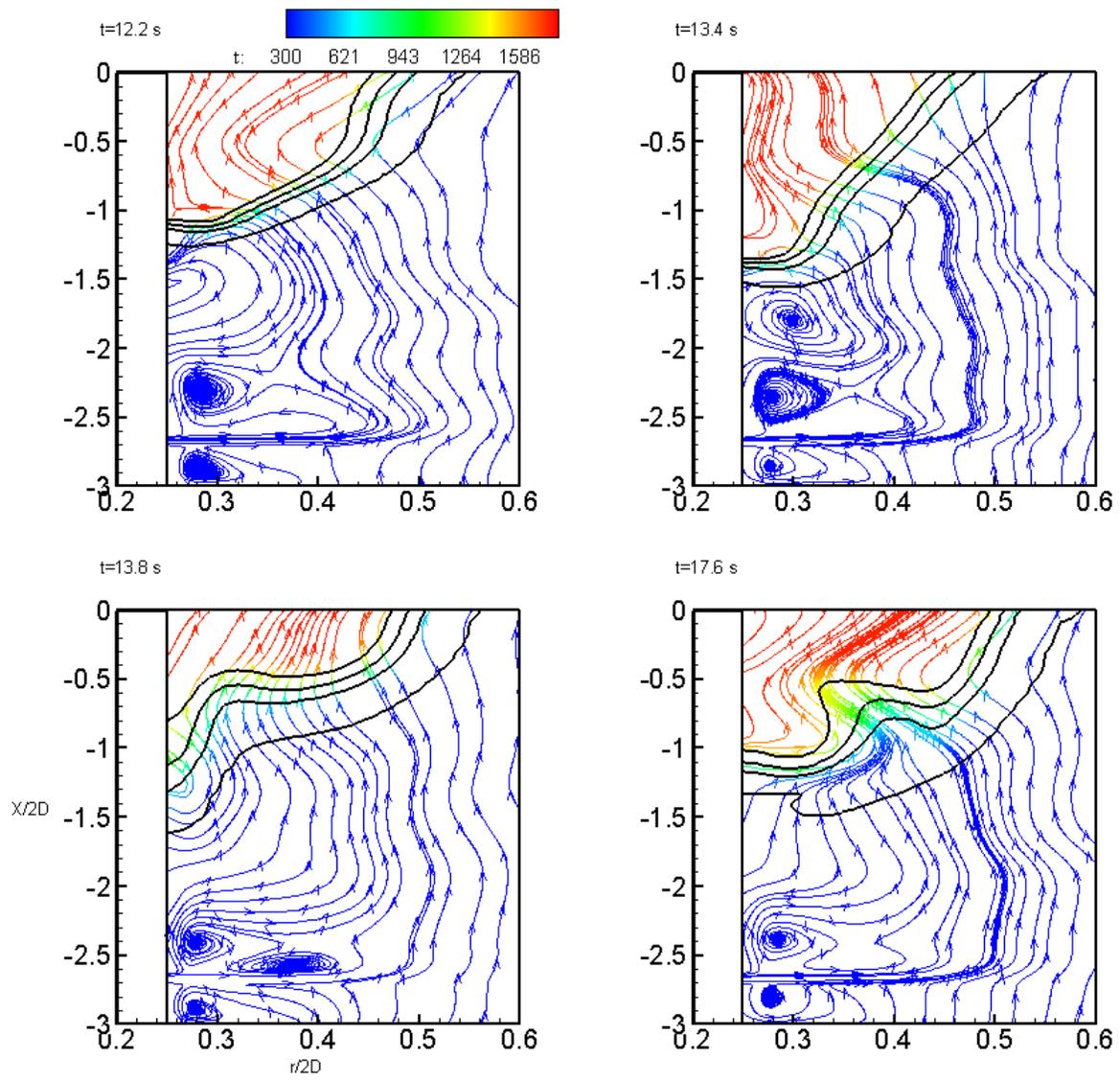

Figure 8

(a)

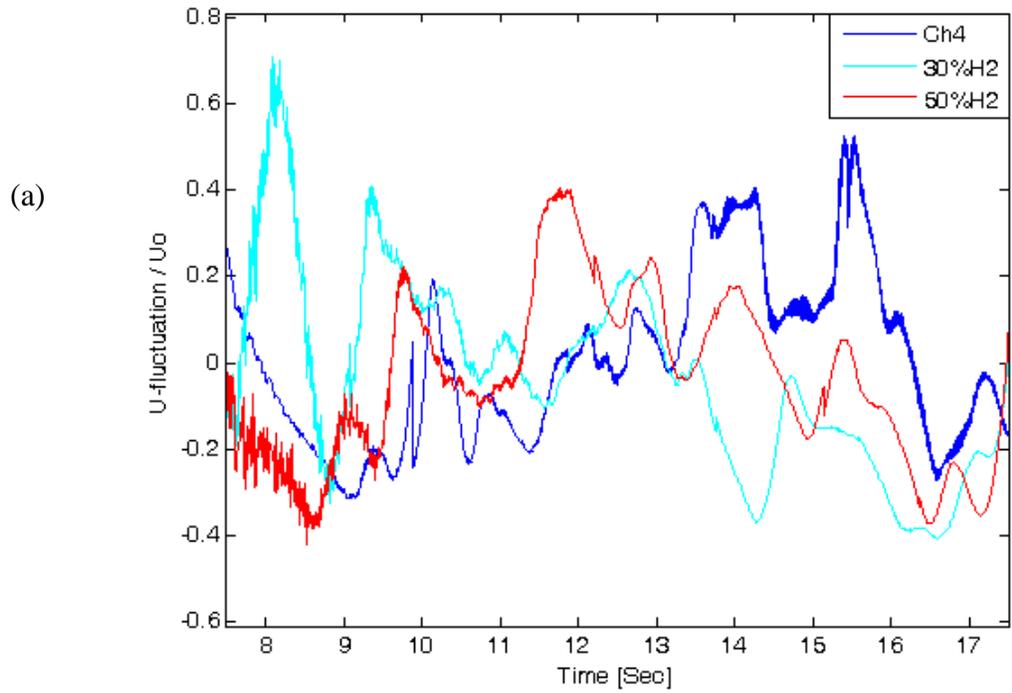

(b)

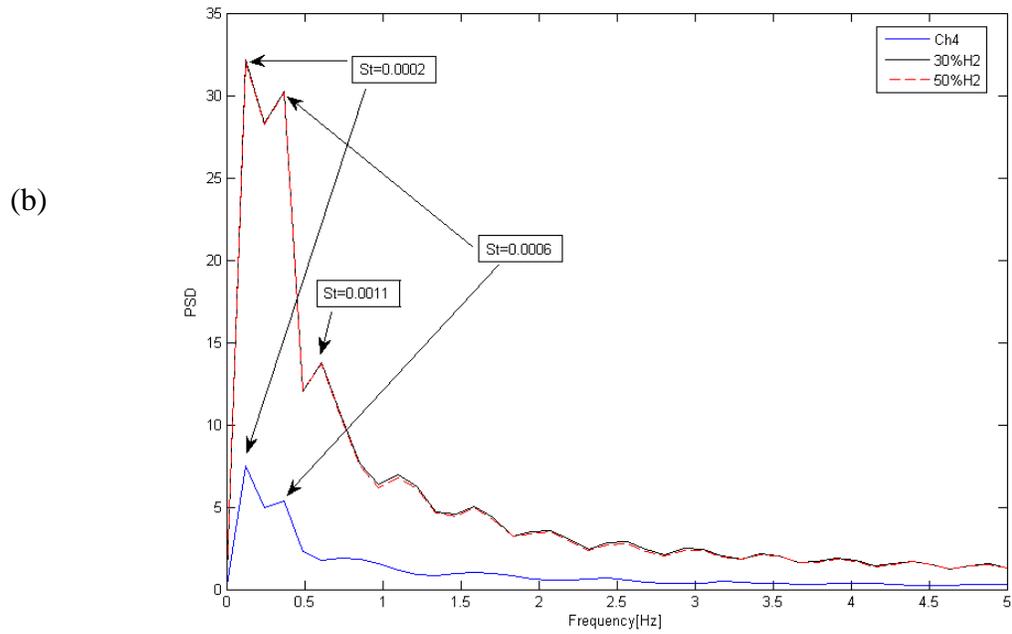

Figure 9

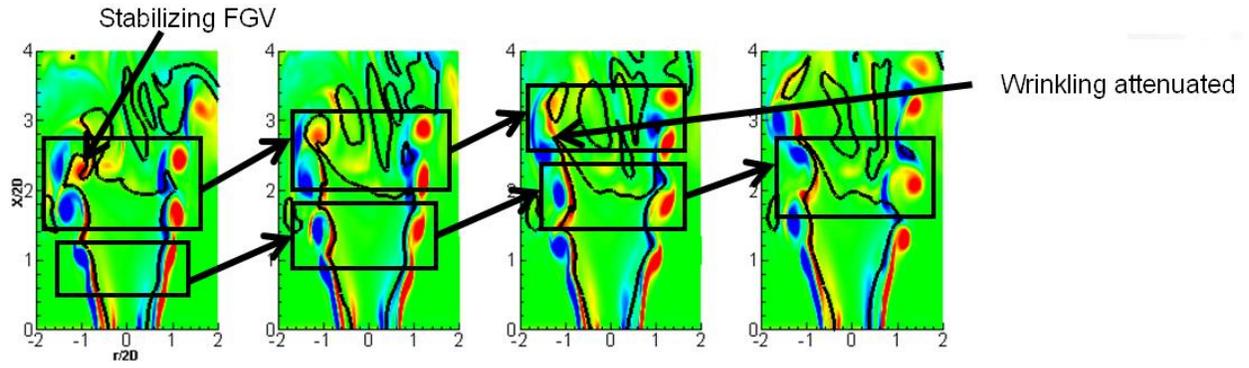

Figure 10

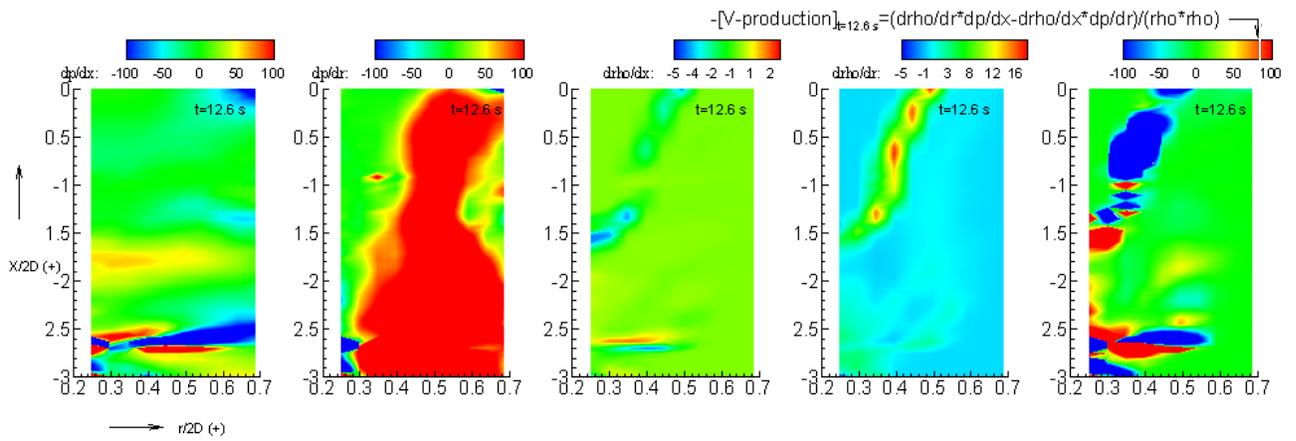

Figure 11

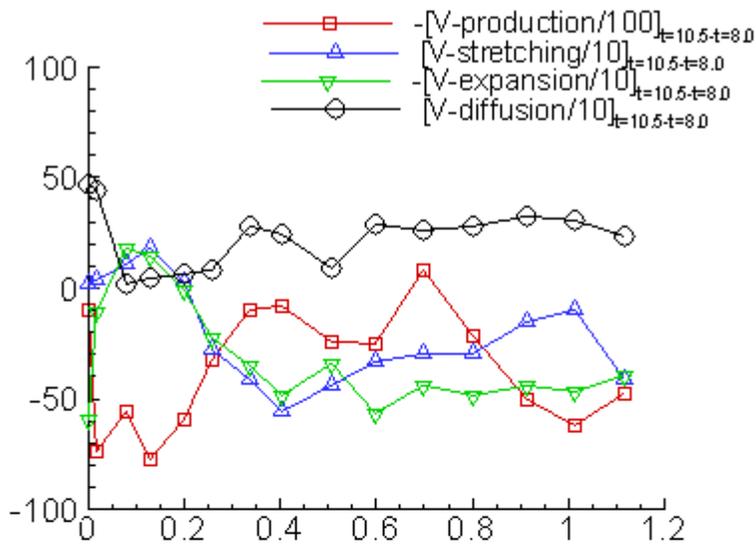

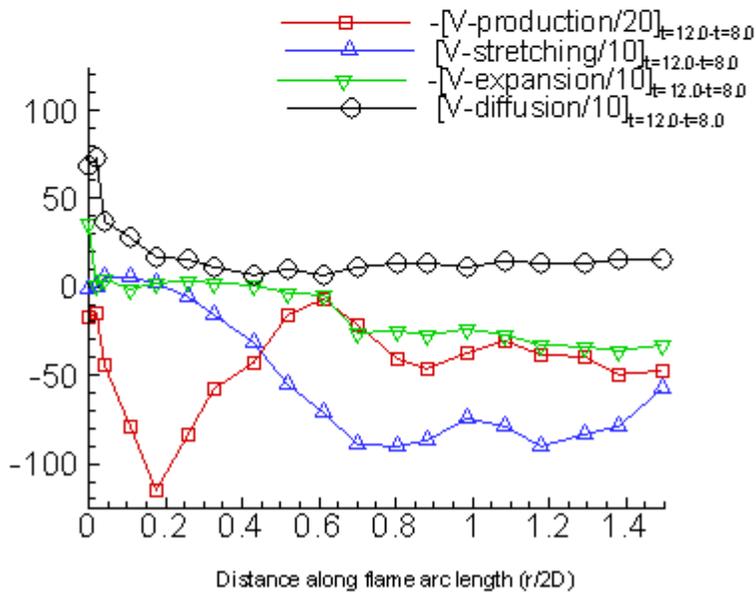

Figure 12